\documentclass[11pt]{article}

\usepackage{setspace}
\usepackage[a4paper,margin=25mm]{geometry}
\usepackage[american]{babel}
\usepackage[T1]{fontenc}
\usepackage{times}
\usepackage[bf,figurename=Fig.,labelsep=period]{caption}
\usepackage{apacite}
\usepackage{amsmath}
\usepackage{algpseudocode}
\usepackage{algorithm}
\usepackage{amssymb}
\usepackage{fancyhdr}
\usepackage{times}

\renewcommand{\APACjournalVolNumPages}[4]{%
  \Bem{#1}
  \ifx\@empty#2\@empty
  \else
    \unskip, \textbf{#2}
  \fi
  \ifx\@empty#3\@empty
  \else
    \unskip({#3})
  \fi
  \ifx\@empty#4\@empty
  \else
    \unskip, {#4}
  \fi
}

\usepackage{amsthm}
\usepackage{epsfig}

\newtheoremstyle{citedtheorem}%
  {3pt}
  {3pt}
  {\itshape}
  {}
  {\bfseries}
  {.}
  {.5em}
  {\thmname{#1} \thmnumber{#2} \thmnote{\normalfont#3}}
\newtheoremstyle{citeddefinition}%
  {3pt}
  {3pt}
  {}
  {}
  {\bfseries}
  {.}
  {.5em}
  {\thmname{#1} \thmnumber{#2} \thmnote{\normalfont#3}}
\theoremstyle{citedtheorem}
\newtheorem{theorem}{Theorem}
\theoremstyle{citeddefinition}

\newtheorem{lemma}{Lemma}

\usepackage{secdot}
\usepackage[compact]{titlesec}
\titleformat{\subsection}
{\normalfont\normalsize}{\thesubsection}{1pt}{~}
\titleformat{\subsubsection}
{\normalfont\normalsize}{\thesubsubsection}{1pt}{~}
\titlespacing{\subsection}{0pt}{0.4cm}{0pt}
\titlespacing{\subsubsection}{0pt}{0.4cm}{0pt}
\titleformat{\section}{\bf}{\thesection}{1pt}{.~}

\algnewcommand\algorithmicforeach{\textbf{foreach}}
\algdef{S}[FOR]{ForEach}[1]{\algorithmicforeach\ #1\ \algorithmicdo}

\algnewcommand{\algorithmicand}{\textbf{and }}
\algnewcommand{\AND}{\algorithmicand}

\algnewcommand{\algorithmicor}{\textbf{or }}
\algnewcommand{\OR}{\algorithmicor}

\algnewcommand{\algorithmicnot}{\textbf{not }}
\algnewcommand{\NOT}{\algorithmicnot}

\newcommand{\Break}{\State \textbf{break} }

\title{\bf Panconnectivity Algorithm for Eisenstein-Jacobi Networks}

\author{
{\bfseries Mohammad Awadh$^{1}$, Zaid Hussain$^{2,*}$ and Hesham Almansouri$^3$}\\
$^{1,2}$Computer Science Department, Kuwait University, Kuwait\\
$^3$Kuwait Institute for Scientific Research, Kuwait\\
*Corresponding author: zhussain@cs.ku.edu.kw
}

\begin{document}

\date{}

\maketitle

\begin{abstract}
The cycles in an interconnection network are one of the communication types that are considered as a factor to measure the efficiency and reliability of the networks' topology. The network is said to be panconnected if there are cycles of length $l$ between two nodes $u$ and $v$, for all $l = d(u, v), d(u, v) +1, d(u, v) +2, \dots, n-1$ where $d(u, v)$ is the shortest distance between $u$ and $v$ in a given network, and $n$ is the total number of nodes in the network. In this paper, we propose an algorithm that generates and proves the panconnectivity of Eisenstein-Jacobi networks by constructing all cycles between any two nodes in the network of length $l$ such that $3 \leq l < n$. The correctness of the proposed algorithm is given with the time complexity $O(n^4)$.
\end{abstract}

\vspace{1em}


\vspace{1cm}

\section{Introduction}\label{sec:introduction}
Parallel computing is expedient for executing independent processes concurrently, which elevates the performance and efficiency of computer tasks. Multiple computer processors are required to perform parallel processes and can be interconnected in a structured topology, which is called an interconnection network. There are several eminent interconnection networks such as $k$-ary $n$-cube \shortcite{dally1990performance}, Torus \shortcite{dally1986torus}, Hypercube \shortcite{hayes1989hypercube} and several other networks \shortcite{bhuyan1984generalized}\shortcite{preparata1981cube}\shortcite{yang2005locally}\shortcite{della1987recursively}\shortcite{flahive2009topology}\shortcite{martinez2008modelingGaussian}.

Communication between processors is vital to attain reliable interconnection network. For example, algorithms such as shortest paths, disjoint paths, broadcasting, and routing have significant impact on the efficiency of an interconnection network. To understand the communication algorithms, interconnection networks are represented as graphs. Variety of efficient algorithms based on paths and cycles were proposed to solve various algebraic and graph problems, some can be found in \shortcite{akl1997parallel}\shortcite{leighton2014introduction}. Embedding paths and cycles of various lengths in interconnection networks is important to determine whether the network topology is suitable for an application in which mapping paths and cycles are necessary. In graph theory, the problem of panconnectivity receives more attention in \shortcite{chang2004panconnectivity}\shortcite{ma2006panconnectivity}\shortcite{sheng1999panconnectivity}\shortcite{hsieh2007panconnectivity}\shortcite{park2008panconnectivity}\shortcite{chen2014panconnectivity}\shortcite{park2007panconnectivity}\shortcite{yuan2013panconnectivity}\shortcite{heidari2019johnson}\shortcite{chen2019r}\shortcite{xu2019fault} to find different paths lengths in interconnection networks. A graph is panconnected if for any source node $S$ and destination node $D$ there exist paths of all different lengths starting with the shortest path between $S$ and $D$ ending with a path of length $N-1$, where $N$ is the number of nodes in the graph \shortcite{alavi1975panconnected}. The panconnectivity of a graph have distinguishing applications in the interconnection networks field. A network can be measured for its reliability by determining its panconnectivity. Furtherly, a panconnected network is capable of embeddability of other networks of special advantages \shortcite{fan2005optimal}\shortcite{fan2007optimal}\shortcite{fang2007bipanconnectivity}\shortcite{lin2011panconnectivity}. 

Eisenstein-Jacoby (EJ) networks proposed in \shortcite{flahive2009topology} as an efficient interconnection topology, which are defined based on Eisenstein-Jacoby integers \shortcite{huber1994codes} and can be modeled on planar graphs. EJ network is a symmetric network of degree 6 and is a generalization of hexagonal networks developed earlier in \shortcite{chen1990addressing} \shortcite{dolter1991performance}. Hexagonal networks have several applications other than interconnection networks, where they are used in cellular networks \shortcite{nocetti2002addressing}, geographical mapping \shortcite{xiao2007subdivision}, and computer graphics \shortcite{lester1984computer}.

In this paper, we present a proof of the panconnectivity of EJ networks and propose an algorithm for finding panconnectivity paths/cycles. Throughout this paper, we use network and graph, node and vertex, and link and edge interchangeably. The paper is structured as follows. Section \ref{sec:background} defines the terminologies and reviews the Eisenstein-Jacobi network and the panconnectivity with its family. The algorithm that shows the panconnectivity in Eisenstein-Jacobi network with its complexity is discussed in Section \ref{sec:Panconnectivity-Algorithm}. 
Finally, the work done in this paper is concluded in Section \ref{sec:conclusion}.

\section{Background\label{sec:background}}
This section reviews the general graph theory notations and terminologies used in this work. Further, it describes the pancyclic, the Hamiltonian connected, and the panconnectivity of a network. In addition, it reviews the topological properties of Eisenstein-Jacobi networks. Finally, it discusses the related work on panconnectivity for some interconnection networks.

\subsection{General Notations and Terminologies\label{sec:terminologies}}
The set of integers is denoted by $\mathbb{Z} = \{\dots, -3, -2, -1, 0, 1, 2, 3,\dots\}$. A graph denoted $G(V,E)$, or \textit{undirected graph}, consists of a set of \textit{vertices} (nodes) denoted $V$ and a set of \textit{edges} (links) denoted $E$. A \textit{directed graph} is similar to the undirected graph except that its edges are directed edges, called \textit{arcs}. In undirected graph $G$, an edge between $u$ and $v$ can be represented as $(u,v)$ or $(v,u)$ where $u,v, \in V$. Whereas, in a directed graph, an edge (arc) from $u$ to $v$ is represented as $(u,v)$ and an edge (arc) from $v$ to $u$ is represented as $(v,u)$. That is, $(u, v)$ is an outgoing arc from $u$ and incoming to $v$. A \textit{path} $P$ of length $|P|$ from $u$ to $v$ in graph $G$ is a sequence of edges that connects the intermediate nodes between two end nodes, $u$ and $v$, and it does not contain a cycle. A \textit{cycle} in graph $G$ is a path where it has at least one node visited twice. A \textit{Euler path} is a path that crosses every edge in the graph $G$ exactly once. Further, a Euler path that starts and ends at the same node is called a \textit{Euler cycle}. In Euler path, it is possible to pass through a node more than once. A \textit{Hamiltonian path} is a path that passes through all the nodes in the graph $G$ exactly once. In addition, a Hamiltonian path that starts and ends at the same node is called a \textit{Hamiltonian cycle}. Hence, when the Hamiltonian path ends at the starting node then we obtain a Hamiltonian cycle. Further, if there is a Hamiltonian path between every two nodes in the graph $G$ then the graph $G$ is called \textit{Hamiltonian-connected}.

Let $G$ be a graph consisting of $n$ nodes that models an interconnection network. Then, $G$ is called \textit{pancyclic} if it contains all possible cycles of length $l$ where $3 \leq l \leq n$ \shortcite{bondy1971pancyclic}. Since Hamiltonian graphs have a cycle of maximum possible length, then the pancyclic graphs are considered as a generalization of Hamiltonian graphs. There are two types of pancyclic graphs which are node-pancyclic and edge-pancyclic \shortcite{randerath1999vertex}. In the former, a graph $G(V,E)$ with $n$ nodes is said to be node-pancyclic if each node $v \in V$ is contained in a cycle of length $l$ where $3 \leq l \leq n$. Similarly, in the later, a graph is edge-pancyclic if each edge $e \in E$ is in a cycle of length $l$ for $l = 3, 4, 5, \dots, n$. $G$ is said to be \textit{panconnected} \shortcite{alavi1975panconnected} if for all nodes $u$ and $v$ there are paths from $u$ to $v$ of length $l$, for all $l = D(u, v), D(u, v) + 1, D(u, v) + 2, \dots, n$, where $D(u, v)$ is the shortest distance between nodes $u$ and $v$. A graph that is panconnected is also \textit{pancyclic}. To illustrate, let $s$ be the source node and $d$ be the destination node in a given panconnected graph, such that $s$ and $d$ are adjacent. The set of panconnectivity paths $P$ contains paths from the shortest path length between $s$ and $d$ to $n-1$. For each path in $P$ excluding the path of length 1, we connect the nodes $d$ and $s$ to form a cycle. Then, all the pancycles can be obtained from length $3$ to $N$. As described in \shortcite{chang2004panconnectivity}, we have Panconnected $\subset$ Pancyclic $\cap$ Hamiltonian-Connected $\subset$ Hamiltonian graphs. Thus, the panconnected graphs are special case of Hamiltonian-Connected graphs, which leads to the existence of a Hamiltonian path in the graph.

\subsection{Eisenstein-Jacobi Networks\label{sec:Eisenstein-Jacobi-Networks}}
The Eisenstein-Jacobi network has been introduced in \shortcite{flahive2009topology}\shortcite{martinez2008modeling} as an efficient topology for interconnection networks. This topology can be modeled on planar graphs. The idea behind Eisenstein-Jacobi network is coming from the quotient ring based on the Eisenstein-Jacobi integers \shortcite{huber1994codes}.

\begin{figure}[!ht]
        \centering
        \includegraphics[scale=0.75]{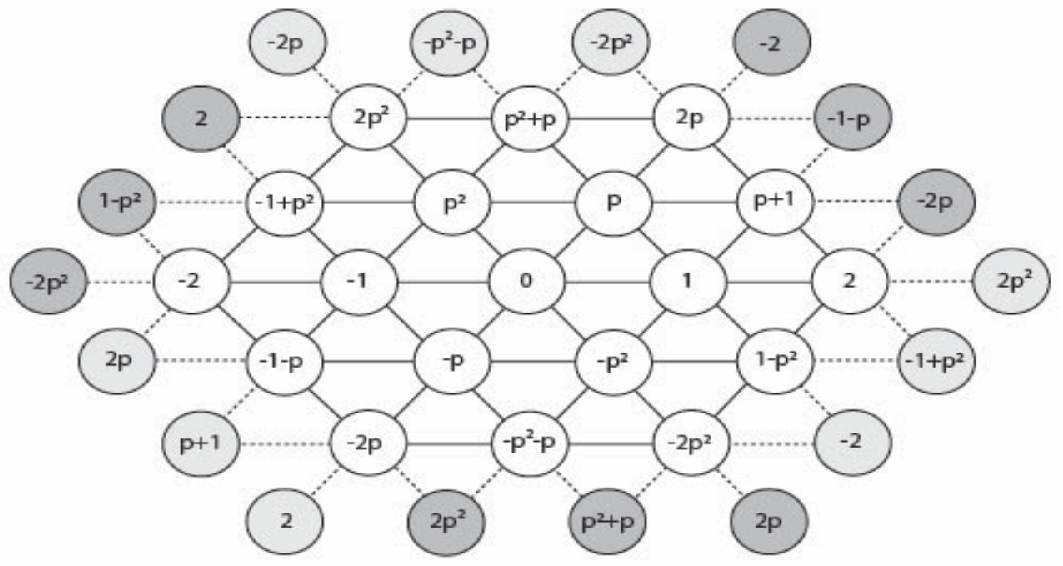}
		\caption{EJ network generated by $\alpha = 2+3\rho$.}
		\label{EJ23}
\end{figure}

The Eisenstein-Jacobi integer is defined as $\mathbb{Z}[\rho] = \{ x+y\rho \mid x,y \in \mathbb{Z}\}$ where $\rho = (1+i\sqrt{3}) / 2$, $i = \sqrt{-1}$, and $\rho^2 = -1 + \rho$. The Eisenstein-Jacobi network is generated by $0 \neq \alpha = a + b\rho \in \mathbb{Z}[\rho]$ such that $0 \leq a \leq b$ and its graph is denoted as $EJ_{\alpha}(V,E)$, where $V = \mathbb{Z}[\rho]_{\alpha}$ is the node set and $E = \{ (\beta, \gamma) \in V \times V \mid (\beta - \gamma) \equiv \pm 1, \pm \rho, \pm \rho^2 \ mod \ \alpha \}$ is the edge set. Eisenstein-Jacobi networks are symmetric 6-regular networks, i.e., the degree of each node is six. Their total number of nodes is $N(\alpha) = a^2 + b^2 + ab$, which is equal to the number of residue classes in Eisenstein-Jacobi integer modulo $\alpha \neq 0$, where each node is labelled as $x+y\rho \ mod \ \alpha$. The distance between two nodes $\beta$ and $\gamma$ is defined as:
\begin{equation}
 D(\beta, \gamma) =
min\{|x| + |y| + |x| \mid (\beta - \gamma) \equiv x + y\rho + z\rho^2 \ (mod \ \alpha)\} 
\end{equation}
When the distance is equal to $\pm 1$, $\pm \rho$, or $\pm \rho^2$ then the two nodes are said to be neighbours. The diameter of the network is computed as $k = \lfloor (a+2b)/3 \rfloor$. The distribution of the nodes on the network shows the number of nodes at certain distance. Given $t = (a+b)/2$ and a distance $s$, where $s = 0, 1, 2, \dots, k$, the number of nodes at distance $s$ in $EJ_\alpha$ is denoted as $W(s)$ \shortcite{flahive2009topology}. The following describes the distance distribution of the network.
\begin{eqnarray}
W(s) = \left\{
{\begin{array}{*{20}{l}}
1&{if \ s = 0}\\
{6s}&{if \ 1 \le s < t}\\
{18(k - s)}&{if \ t < s < k}\\
2&{if \ b \equiv a \ (mod \ 3) \ and \ s = k}\\
0&{if \ s > k}\\
{N(\alpha ) - R}&{if \ s = t}
\end{array}} \right.
\end{eqnarray}
where $R = \sum_{s = 0,s \ne t}^m W(s)$. Eisenstein-Jacobi network is called dense when $b = a + 1$, i.e., it has a maximum number
of nodes for a given diameter. Figure \ref{EJ23} depicts the Eisenstein-Jacobi network generated by $\alpha = 2+3\rho$.

From Figure \ref{EJ23}, it is obvious that we have two types of edges, regular and wraparound edges. The regular edges are the edges that connect two nodes within the network and represented by solid lines. Whereas, the wraparound edges are the edges that connect two boundary nodes and they are illustrated by dotted lines. The wraparound edges are obtained by using the $mod \ \alpha$ operation. Given a node $u = x + y\rho + z\rho^2$, a neighbour node can be obtained by the adding $\pm1$, $\pm\rho$, or $\pm\rho^2$ to the node and then applying the $mod \ \alpha$ operation.

\subsection{Shortest Path Algorithm\label{sec:Shortest-Path-Algorithm}}
The shortest path algorithm is essential for the work presented in this paper. There are several algorithms for finding the shortest path in certain topological networks or ad-hoc networks. The algorithm used in this work is the Dynamic Source Routing Protocol (DSR), which is used in mobile ad-hoc networks. However, the DSR algorithm will run on a computational basis instead of communicational basis. The main reason for choosing the DSR algorithm is its complexity is low after embedding it in the proposed algorithm compared with other well-known shortest path algorithms like Dijkstra, Bellman Ford, Floy Warshall, ...etc. The DSR algorithm is adopted in the proposed algorithm in Section \ref{sec:Pseudocode} and its complexity is measured in Section \ref{sec:Complexity-of-the-Algorithm}.

The DSR protocol starts by getting the source node ID, destination node ID, request ID, and a route record as arguments. The protocol starts by sending a route request from the source node to all its neighbours. Each node receiving the route request will check the information of the route request. First, this intermediate node will return the route record and adding its ID to it if its ID is equal to the destination ID. Second, the intermediate node will ignore the route request if the source ID and request ID are matching the previous request, or if the node ID exists in the route record. Otherwise, the intermediate node rebroadcast the route request after adding itself to the route record. However, in the case of the computational algorithm and since the graph topology is previously known, then the algorithm will take the graph and its diameter instead of the request ID and the route record. Then, the DSR algorithm starts by sending a route request from the source node to its neighbours and iteratively repeat the steps for every node reached by sending the route request to its neighbours and marking each visited node with a flag. A node that has a visited flag will not be reached again. If the destination is reached, the path to the destination will be returned.

\section{Panconnectivity Algorithm\label{sec:Panconnectivity-Algorithm}}
\subsection{Chain Algorithm\label{sec:Chain-Algorithm}}
\begin{figure}[!ht]
        \centering
        \includegraphics[scale=0.65]{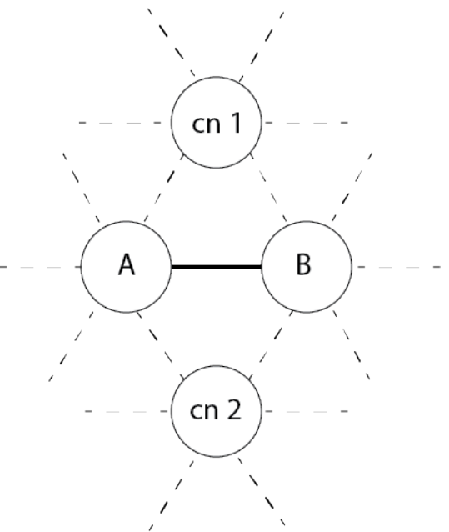}
		\caption{Common neighbours ($cn1$ and $cn2$) in EJ network between two adjacent nodes $A$ and $B$.}
		\label{commonNeighbour}
\end{figure}

In perspective, the idea came from the connection between any two adjacent nodes and their common neighbours. In Eisenstein-Jacobi network, for each two adjacent nodes there are two common neighbours, $cn1$ and $cn2$, as shown in Figure \ref{commonNeighbour}. Since the path from node $s$ to node $d$ is a sequence of adjacent nodes, each time the algorithm increases its length by one node or more by checking the first two adjacent nodes. If their common neighbours are not exist in the path, then the algorithm adds one of them to the path. On the other hand, if they exist in the path then the algorithm will pass over to the next two adjacent nodes in order to achieve the panconnectivity of the network. Finally, when the algorithm reaches the last two adjacent nodes in a path and there are no more common neighbors, then it means that this is the longest path in the network between $s$ and $d$. Also, the length of the path is $n$, where $n$ is the total number of nodes in the network.

\begin{figure}[ht]
        \centering
        \includegraphics[scale=0.5]{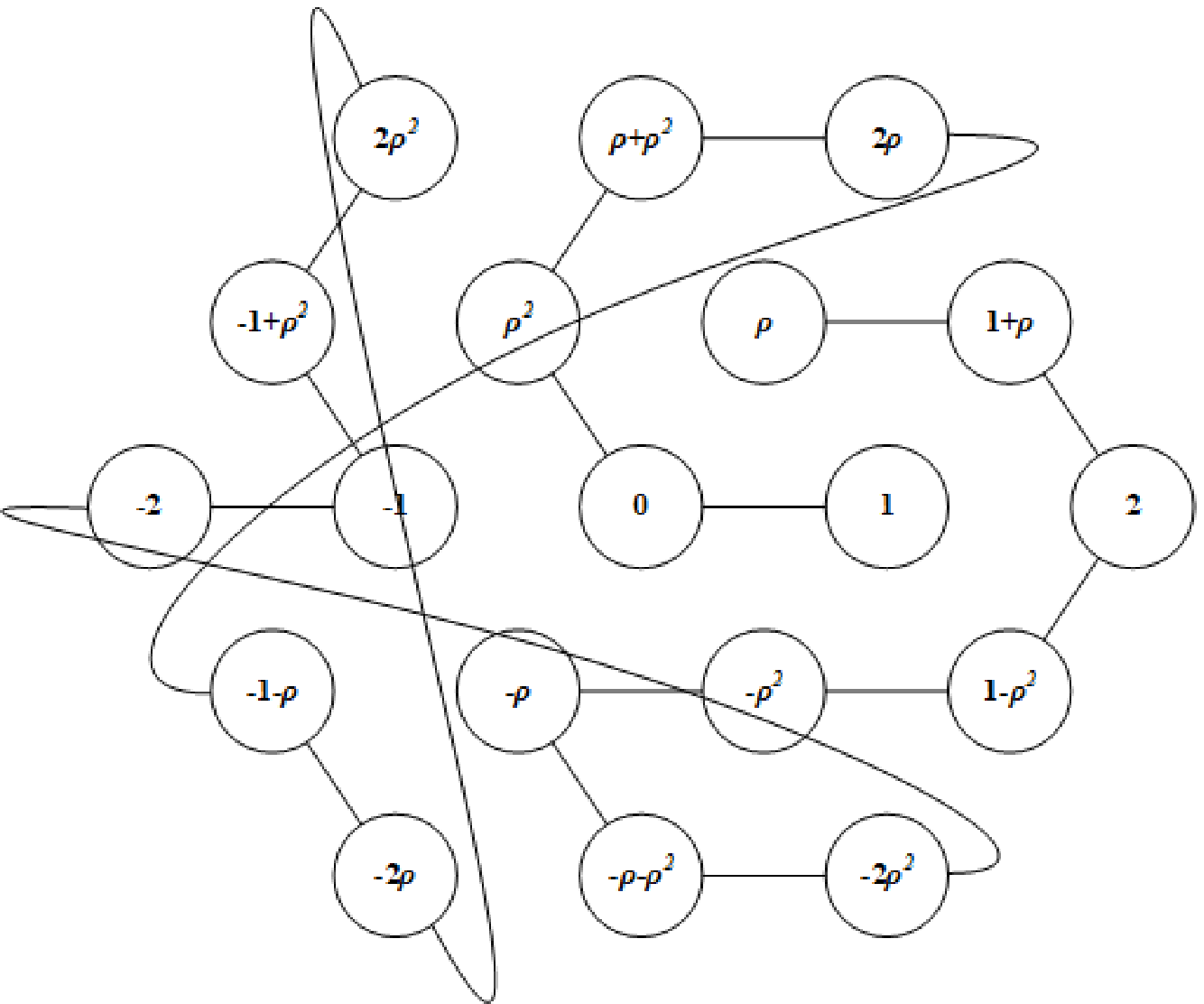}
		\caption{Longest path of length 18 between nodes $\rho$ and 1 in EJ with $\alpha = 2+3\rho$.}
		\label{EJ23_longestPath}
\end{figure}

First of all, every two nodes must have two common neighbors, which is proved in Lemma \ref{lemma2} in Section \ref{sec:Correctness-of-the-Algorithm}. Chain Algorithm, which is explained later, proves the existance of and solves the panconnectivity in Eisenstein-Jacobi networks. The algorithm requires three parameters, which are the source and destination nodes, and the shortest distance between them. The first node that the path builds on is the source node, denoted $s$. The last node that the path ends with is the destination node, denoted $d$. Therefore, in each iteration a new path is generated that is longer than the previous path by 1. Initially the length is the number of hops in the shortest path. For instance, consider Eisenstein-Jacobi network generated with $\alpha = 2+3\rho$, where the source node $s=\rho$ and the destination node $d=1$. Further, if Chain Algorithm starts to find another path leading from $s$ all the way to $d$, then the algorithm looks for the two common neighbours, which are $cn1 = \rho+1$ and $cn2 = 0$, that are connected to the two main nodes ($s$ and $d$). Since the two common neighbours do not exist in the path, then the algorithm adds one of them to the path. Assume $cn2$ is added, then the new path is ($s$, $cn2$, $d$), and its length is 2. The algorithm repeats the same steps between $cn2$ and $d$, and so on, until the longest path between $s$ and $d$ is constructed. Since $s$ and $d$ are adjacent, then this path is Hamiltonian Path as depicted in Figure \ref{EJ23_longestPath}. Table \ref{EJ23Paths} shows all possible paths lengths between $s=\rho$ and node $d=1$. The reason to choose two adjacent nodes is to prove that algorithm can also solve pancyclic by adding the source node at the end of each path starting from length 2. However, Chain Algorithm can construct both the Hamiltonian path and the Hamiltonian cycle.

\begin{table}[!h]
\centering
\caption{All possible paths length between nodes $\rho$ and 1 in EJ with $\alpha = 2+3\rho$.}
\begin{tabular}{|c|c|}
\hline
Length & Path (sequence of nodes) \\ \hline
1 & $\rho, 1$ \\ \hline
2	& $\rho, 0, 1$ \\ \hline
3	& $\rho, \rho^2, 0, 1$ \\ \hline
4	& $\rho, \rho+\rho^2, \rho^2, 0, 1$ \\ \hline
5	& $\rho, 2\rho, \rho+\rho^2, \rho^2, 0, 1$ \\ \hline
6	& $\rho, \rho+1, 2\rho, \rho+\rho^2, \rho^2, 0, 1$ \\ \hline
7	& $\rho, \rho+1, -1-\ \rho, 2\rho, \rho+\rho^2, \rho^2, 0, 1$ \\ \hline
8	& $\rho, \rho+1, -2\rho, -1-\rho, 2\rho, \rho+\rho^2, \rho^2, 0, 1$ \\ \hline
9	& $\rho, \rho+1, 2, -2\rho, -1-\rho, 2\rho, \rho+\rho^2, \rho^2, 0, 1$ \\ \hline
10	& $\rho, \rho+1, 2, 2\rho^2, -2\rho, -1-\rho, 2\rho, \rho+\rho^2, \rho^2, 0, 1$ \\ \hline
11	& $\rho, \rho+1, 2, -1+\rho^2, 2\rho^2, -2\rho, -1-\rho, 2\rho, \rho+\rho^2, \rho^2, 0, 1$ \\ \hline
12	& $\rho, \rho+1, 2, 1-\rho^2, -1+\rho^2, 2\rho^2, -2\rho, -1-\rho, 2\rho, \rho+\rho^2, \rho^2, 0, 1$ \\ \hline
13	& $\rho, \rho+1, 2, 1-\rho^2, -2, -1+\rho^2, 2\rho^2, -2\rho, -1-\rho, 2\rho, \rho+\rho^2, \rho^2, 0, 1$ \\ \hline
14	& $\rho, \rho+1, 2, 1-\rho^2, -2\rho^2, -2, -1+\rho^2, 2\rho^2, -2\rho, -1-\rho, 2\rho, \rho+\rho^2, \rho^2, 0, 1$ \\ \hline
15	& $\rho, \rho+1, 2, 1-\rho^2, -\rho^2, -2\rho^2, -2, -1+\rho^2, 2\rho^2, -2\rho$, $-1-\rho, 2\rho, \rho+\rho^2, \rho^2, 0, 1$ \\ \hline
16	& $\rho, \rho+1, 2, 1-\rho^2, -\rho^2, -\rho-\rho^2, -2\rho^2$, \\ & $-2, -1+\rho^2, 2\rho^2, -2\rho, -1-\rho, 2\rho, \rho+\rho^2, \rho^2, 0, 1$ \\ \hline
17	& $\rho, \rho+1, 2, 1-\rho^2, -\rho^2, -\rho, -\rho-\rho^2$, \\ & $-2\rho^2, -2, -1+\rho^2, 2\rho^2, -2\rho, -1-\rho, 2\rho, \rho+\rho^2, \rho^2, 0, 1$ \\ \hline
18	& $\rho, \rho+1, 2, 1-\rho^2, -\rho^2, -\rho, -\rho-\rho^2, -2\rho^2$, \\ & $ -2, -1, -1+\rho2, 2\rho^2, -2\rho, -1-\rho, 2\rho, \rho+\rho^2, \rho^2, 0, 1$ \\ \hline
\end{tabular}
\label{EJ23Paths}
\end{table}

\subsection{The Pseudocode\label{sec:Pseudocode}}
This section discusses the detailed description of each algorithm related to Chain Algorithm. For simplification, the whole algorithm is broken down into several sub algorithms. The main results achieved by Chain Algorithm are the panconnectivity of the Eisenstein-Jacobi network.

The node structure is used in each node to define their attributes. Each node has four attributes, which are the \textit{ID}, \textit{Path}, \textit{Visited}, and \textit{hops}. The \textit{ID} attribute is a unique identifier for the node, which is the node address. The \textit{Path} is a sequence of adjacent nodes from the source node to the current node. The \textit{Visited} attribute determines whether the function has been called or not. Finally, the \textit{hops} attribute is an integer that records the distance from the source node to the current node. The node structure will be an essential part of the algorithms described furtherly in this section.

Algorithm \ref{algorithm1} takes the network $G$ and its diameter $k$ as arguments and returns $k$ nodes. Each returned node has different shortest distance from the source node. The algorithm starts by communicating with the neighbours of the source node, and iteratively recommunicate until all nodes of different distances are reached and saved in the list. Finally, a list of nodes is returned to be used in another algorithm.

\begin{algorithm}[!h]
\caption{BroadcastList}
\begin{algorithmic}[1]
\renewcommand{\algorithmicrequire}{\textbf{Input:}}
\renewcommand{\algorithmicensure}{\textbf{Output:}}
\Require Network $G$, Diameter $d$
\Ensure Broadcast List
\State queue $\leftarrow$ []
\State // add node 0 in queue as source node
\State queue.push($G$.node.ID('0'))
\State broadcastList $\leftarrow$ null
\State counter $\leftarrow$ 1

\While{queue is not empty \AND counter $\leq$ $d$}
    \State parentNode $\leftarrow$ queue.pop()
	\State neighbours $\leftarrow$ $G$.neighbours(parentNode.ID)
	\ForEach{node in neighbours}
        \State node.hops $\leftarrow$ parentNode.hops + 1
        \State node.Path $\leftarrow$ parentNode.Path.append(node.ID)
            \If{\NOT node.Visited \AND node.hops = counter}
                \State broadcastList.append(node)
                \State counter $\leftarrow$ counter + 1 
            \EndIf
            \State node.Visited $\leftarrow$ True
             \State queue.push(node) 
	\EndFor
\EndWhile
\If{counter $>$ $d$} 
    \State \Return broadcastList
\Else 
   \State \Return null
\EndIf
\end{algorithmic}
\label{algorithm1}
\end{algorithm}

\begin{algorithm}[!h]
\caption{ShortestPath}
\begin{algorithmic}[1]
\renewcommand{\algorithmicrequire}{\textbf{Input:}}
\renewcommand{\algorithmicensure}{\textbf{Output:}}
\Require Network $G$, Source $s$, Destination $d$
\Ensure Shortest Path
\State queue.push($s$)
\While{queue is not empty}
    \State parentNode $\leftarrow$ queue.pop()
	\State neighbours $\leftarrow$ $G$.neighbours(parentNode.ID)
	\ForEach{node in neighbours}
            \If{\NOT node.Visited}
                \State node.hops $\leftarrow$ parentNode.hops + 1
                \State node.Path $\leftarrow$ parentNode.Path.append(node.ID)
                
                \If{node.ID = $d$.ID} 
                    \State \Return node.Path
                \EndIf
                \State node.Visited $\leftarrow$ True
                \State queue.push(node) 
            \EndIf
	\EndFor
\EndWhile
\end{algorithmic}
\label{algorithm2}
\end{algorithm}

\begin{algorithm}[!h]
\caption{ChainAlgorithm}
\begin{algorithmic}[1]
\renewcommand{\algorithmicrequire}{\textbf{Input:}}
\renewcommand{\algorithmicensure}{\textbf{Output:}}
\Require Network $G$, Number of Nodes $n$, Path $P$
\Ensure All possible paths between $s$ and $d$
\State distance $\leftarrow$ length($P$) $ - \ 1$
\State paths $\leftarrow$ []
\State paths.append($P$)
\While{distance $< n$}
    \State start $\leftarrow$ 0 
    \State end $\leftarrow$ length($P$) $ - \ 1$
    \State found $\leftarrow$ False 
    \While{start $<$ end \AND \NOT found}
        \State commonNeighbours $\leftarrow G$.commonNeighbours(path[start], path[start+1])
        \If{commonNeighbours[0] \NOT in path} 
            \State paths.append(start+1, commonNeighbours[0])
            \State distance $\leftarrow$ distance + 1
            \State found $\leftarrow$ True 
            \Break
        \ElsIf{commonNeighbours[1] \NOT in path} 
            \State paths.append(start+1, commonNeighbours[1])
            \State distance $\leftarrow$ distance + 1
            \State found $\leftarrow$ True
            \Break
        \Else
            \State start $\leftarrow$ start + 1
        \EndIf 
	\EndWhile
\EndWhile
\If{\NOT found}
    \Return null
\Else{ paths.append($P$)}
\EndIf
\State \Return paths
\end{algorithmic}
\label{algorithm3}
\end{algorithm}

\begin{algorithm}[!h]
\caption{PanconnectivityList}
\begin{algorithmic}[1]
\renewcommand{\algorithmicrequire}{\textbf{Input:}}
\renewcommand{\algorithmicensure}{\textbf{Output:}}
\Require Network $G$
\Ensure hashList have every paths between any two nodes
\State hashList $\leftarrow$ []
\State NL $\leftarrow G$.nodes() // list of nodes in the network
\For{i $\leftarrow$ 0; i $< G$.numberNodes(); i++}
    \For{j $\leftarrow$ i+1; j $< G$.numberNodes(); j++}
        \State tempList $\leftarrow$ ChainAlgorithm(NL[i].ID, NL[j].ID, $G$.numberNodes(), shortestPath($G$, $G$.diameter, NL[i].ID, NL[j].ID))
        \If{tempList is \NOT empty}
            \State hashList[NL[i].ID, NL[j].ID] $\leftarrow$ tempList
        \Else
            \State \Return False
        \EndIf
    \EndFor
\EndFor
\State \Return hashList
\end{algorithmic}
\label{algorithm4}
\end{algorithm}

\begin{algorithm}[!h]
\caption{CheckPanconnectivity}
\begin{algorithmic}[1]
\renewcommand{\algorithmicrequire}{\textbf{Input:}}
\renewcommand{\algorithmicensure}{\textbf{Output:}}
\Require Network $G$
\Ensure True / False
\State List $\leftarrow$ broadcastList($G$, $G$.diameter)
\While{List is not empty}
    \If{ChainAlgorithm($G$.node.ID('0'), node.ID, $G$.numberOfNodes, node.Path ) = null} 
        \State \Return False
    \EndIf
\EndWhile
\State \Return True
\end{algorithmic}
\label{algorithm5}
\end{algorithm}

Algorithm \ref{algorithm2} takes three parameters, which are the network $G$, source node $s$, and the destination node $d$. The algorithm starts with the source node by communicating path discovery message to its neighbours and checks whether one of them is the destination node then it returns the shortest path. If the destination node is not found, then it continues communicating till it reach the destination node.

Algorithm \ref{algorithm3} is the main algorithm that returns all paths between the starting and ending nodes ($s$ and $d$, respectively) that have lengths between the length of the shortest path to $n-1$, where $n$ is the number of nodes in the network. The algorithm takes the network $G$, number of nodes $n$, and the shortest path $P$ between $s$ and $d$ as arguments. The algorithm starts with the shortest path then increments the distance by adding an unused common neighbour between two adjacent nodes from the previous path. By repeating this step, the algorithm will eventually generate all the paths between $s$ and $d$ of all lengths.

Algorithm \ref{algorithm4} takes the network as an input and retrieves all the paths between all pairs of nodes in the network. The algorithm is based on two embedded loops that run through all the nodes in the network covering all possible pairs of nodes as source and destination. For each pair of nodes $s$ and $d$, Chain Algorithm is called to retrieve all the paths between $s$ and $d$ from the shortest distance to the length of the Hamiltonian path of the network.

Algorithm \ref{algorithm5} takes a network as input and returns boolean output determines whether the network is panconnected. The algorithm starts by getting the list from Algorithm \ref{algorithm1} then it sends a set of nodes containing each node in the list with the source node to Algorithm \ref{algorithm3}. If any pair of nodes returns null then the algorithm will return false. Otherwise, the algorithm will return true.

\subsection{Correctness of the Algorithm\label{sec:Correctness-of-the-Algorithm}}
In order to prove that the algorithm solves the panconnectivity problem in EJ network, the following lemmas are introduced to support the final proof of the algorithm.

\begin{lemma}
\label{lemma1}

Given a shortest path $P$ between a pair of nodes $n_0$ and $n_1$ with length $d$, a similar path is exist between any other pair of nodes in the network other than $n_0$ and $n_1$.
\end{lemma}
\begin{proof}
Let the source node be $s=n_0$, the destination node be $d=n_1$, and a path consisting of a sequence of nodes $P=(n_0, m_1,\ m_2, \dots, m_i, \dots, n_1)$, for $1 \leq i \leq l-2$ where $m_i$ is an intermediate node, is a path between $n_0$ and $n_1$ of length $l$. By the symmetry property of Eisenstein-Jacobi network, any node $u$ can be relocated to be in the position $p = x + y\rho + z\rho^2$ by subtracting its value by $p'$ such that $p = u - p' \ mod \ \alpha$, where $u$ and $p$ are the nodes addresses before and after relocation, respectively. For a given source node $s=n_0$ and a destination node $d=n_1$, relocating the node $n_0$ to be in the position of the node 0 by subtracting itself ($n_0 - n_0$) and all other nodes $m_i$ in path by $n_0$ (i.e., $m^{'}_i = m_i - n_0$). Accordingly, this results a path $P'=(0, {m'}_1, {m'}_2, \dots, {m'}_i, \dots, {n'}_1)$, for $1 \leq i \leq l-2$, of length $l$.
\end{proof}

\begin{lemma}
\label{lemma2}
There are two common neighbours for any pair of adjacent nodes in the EJ network.
\end{lemma}
\begin{proof}
Let $u$ be any node in EJ network and $v$ be the neighbouring node of $u$. If $u=x+y\rho+z\rho^2$, then the set of neighbours of $u$, denoted $N_u$, is as follows:

\begin{eqnarray}
N_{u} = \left\{{
    \begin{array}{*{20}{l}}
        (x+1) + y\rho + z\rho^2 \\
        (x-1) + y\rho + z\rho^2 \\
        x + (y+1)\rho + z\rho^2 \\
        x + (y-1)\rho + z\rho^2 \\
        x + y\rho + (z+1)\rho^2 \\
        x + y\rho + (z-1)\rho^2 \\
    \end{array}} \right\}
\end{eqnarray}

\noindent Since $\rho^2 = \rho - 1$ then $N_{u}$ can be simplified as:

\begin{eqnarray}
N_{u} = \left\{{
    \begin{array}{*{20}{l}}
        x + 1 + y\rho + z\rho - z \\
        x - 1 + y\rho + z\rho - z \\
        x + y\rho + \rho + z\rho - z \\
        x + y\rho - \rho + z\rho - z \\
        x + y\rho + z\rho - z + \rho - 1 \\
        x + y\rho + z\rho - z - \rho + 1 \\
    \end{array}} \right\}
\end{eqnarray}

\noindent Let $v = (x+1) + y\rho + z\rho^2$, then the set of neighbours of $v$, denoted $N_v$, is as follows:

\begin{eqnarray}
N_{v} = \left\{{
    \begin{array}{*{20}{l}}
        x + 2 + y\rho + z\rho - z \\
        x + y\rho + z\rho - z \\
        x + 1 + y\rho + \rho + z\rho - z \\
        x + 1 + y\rho - \rho + z\rho - z \\
        x + y\rho + z\rho - z + \rho \\
        x + 2 + y\rho + z\rho - z - \rho \\
    \end{array}} \right\}
\end{eqnarray}

\noindent Then, $u$ and $v$ have the following common neighbours:

\begin{eqnarray}
N_{u} \cap N_{v} = \left\{{
    \begin{array}{*{20}{l}}
        x + y\rho + \rho + z\rho - z \\
        x + y\rho + z\rho - z - \rho + 1 \\
    \end{array}} \right\}
\end{eqnarray}

\noindent The same steps can be followed to generate the two common neighbours between any two adjacent nodes.
\end{proof}

\begin{lemma}
\label{lemma3}
For any path $P$ of length $|P|$, where $|P| < n-1$, $|P|$ can be increased by 1 after adding node $w \in V-P$.
\end{lemma}
\begin{proof}
Let $G(V,E)$ be a network where $V$ is a set of nodes and $E$ is a set of edges. Let $P$ be a path containing visited nodes in the network such that $P \subseteq V$ and its length is $|P|$. Additionally, let $R = V - P$ be a set of unvisited nodes in the network. Based on Lemma \ref{lemma2}, since $|R| > 0$ then there exist two adjacent nodes $u,v \in P$ that have a common neighbour node $w\in R$ such that $N_u \cap N_v \in R$. Thus, the length of path $P$ can be increased by 1 after adding node $w$ to $P$ and removing it from $R$.
\end{proof}

\begin{lemma}
\label{lemma4}
Algorithms \ref{algorithm1} and \ref{algorithm2} generate the shortest path between any two nodes.
\end{lemma}
\begin{proof}
Let $u$ be a node at distance $d$ from the source node $s$. Since the algorithm communicates with all unvisited neighbours, by Algorithm \ref{algorithm1} lines 7, 8, and 11; and by Algorithm \ref{algorithm2} line 6, then any unvisited neighbour of $u$ will have a distance $d' = d+1$ and it can be reached by the shortest path from $u$. Since the algorithm does not communicate with visited neighbours, then for any visited node at distance $k \in \mathbb{N}^+$ from $u$, the nodes at distance $k' < k$ are already visited.
\end{proof}

\begin{theorem}
\label{theorem1}
EJ network is panconnected.
\end{theorem}
\begin{proof}
Algorithm 4 generates all paths of lengths $l$ in EJ networks, such that $3 \leq l \leq n-1$. By Lemma \ref{lemma4}, the algorithm always generates the shortest path between any two nodes. After generating a shortest path (say of length $h$), by Lemma \ref{lemma3}, the algorithm keeps adding one node in each iteration to find all path lengths from $h$ to $n-1$, where $n$ is the total number of nodes in the network. By repeating the algorithm for all distances from 1 to $k$, where $k$ is the diameter of the network, the algorithm finds all possible shortest paths of lengths $l$ for $l=h$ to $k$. By Lemma \ref{lemma1}, Since the Eisenstein-Jacobi networks are symmetric, then it exempt the algorithm from finding the paths for several sets of sources and destinations of the same shortest path. Consequently, the algorithm generates all paths of lengths $l$, such that $3 \leq l \leq n-1$ in a given Eisenstein-Jacobi network. Thus, EJ networks are panconnected.
\end{proof}

\subsection{Complexity of the Algorithm\label{sec:Complexity-of-the-Algorithm}}
The Chain Algorithm is the core algorithm that solves the panconnectivity problem in the Eisenstein-Jacobi network. The complexities of Algorithms \ref{algorithm1} - \ref{algorithm5} are symbolized as, in respective order, $C_1, C_2, C_3, C_4,$ and $C_5$. Each complexity is calculated separately and then accumulated as a whole algorithm complexity. Starting with node structure, which does not have any loops and have a complexity of $O(1)$. Algorithm \ref{algorithm1} calls two nested loops, where the outer loop has a worst case of the number of nodes, which is $n$, in the network and the inner loop has a worst case of the degree of the Eisenstein-Jacobi network, which is 6. The resulting complexity for the Algorithm \ref{algorithm1} is $C_1=O(n)$. The  Algorithm \ref{algorithm2} is similar to Algorithm \ref{algorithm1} in terms of complexity, where its complexity is $C_2=O(n)$.  Algorithm \ref{algorithm3} has two nested loops, where both of them have the worst case of the number of nodes in the network resulting the complexity $C_3=O(n^2)$. The complexity of the Algorithm \ref{algorithm4} depends on  Algorithm \ref{algorithm3}. The Algorithm \ref{algorithm4} has two nested loops that run through the number of nodes in the network as a worst case and calls the Algorithm \ref{algorithm3} in the inner loop, which results in having the complexity $C_4=O(n^2 C_3)=O(n^4)$. Finally, Algorithm \ref{algorithm5} has a single loop that iterates for length of the diameter of the network $d$. However, inside the single loop the Algorithm \ref{algorithm3} is also called, which will result in the complexity $C_5=O(dC_3)=O(dn^2)$. 

\section{Conclusion\label{sec:conclusion}}
Paths and cycles are two important types of networks that can be used as communication methods where algorithms can be designed with low costs. This motivate us to embed paths and cycles in EJ networks. In this paper, we have investigated the solution of panconnectivity problem in EJ networks. We have shown that the EJ network is panconnected by constructing all possible paths of length $l$ such that $1 \leq l \leq n$, where $n$ is the total number of nodes in a given network. In addition, we have proposed the Chain Algorithm that sequentially constructs all possible paths between any two nodes in Eisenstein-Jacobi networks. The complexity of the proposed algorithm is computer as $O(n^4)$. As for future work, we will extend our work to develop a parallel construction algorithm.

\titleformat{\section}{\bf}{\thesection}{1pt}{}

\bibliographystyle{apacite}
\bibliography{article}

\end{document}